\documentclass[aip,jcp,amsmath,reprint]{revtex4-2}

\usepackage{bbm}
\usepackage{graphicx}
\renewcommand{\vec}[1]{\mathbf{#1}}
\usepackage{hyperref}
\usepackage[capitalize]{cleveref}
\crefname{figure}{Fig.}{Fig.}
\usepackage[T1]{fontenc} 
\usepackage[utf8]{inputenc}
\UseRawInputEncoding

\begin{document}

\title{Newtonian Event-Chain Monte Carlo and Collision Prediction with Polyhedral Particles}

\author{Marco Klement}
\affiliation{Institute for Multiscale Simulation, IZNF, Friedrich-Alexander University Erlangen-N{\"u}rnberg, 91058 Erlangen, Germany}

\author{Sangmin Lee}
\affiliation{Department of Chemical Engineering, University of Michigan, Ann Arbor, MI 48109, USA}
\affiliation{Department of Biochemistry, University of Washington, Seattle, WA 98195, USA}

\author{Joshua A. Anderson}
\affiliation{Department of Chemical Engineering, University of Michigan, Ann Arbor, MI 48109, USA}

\author{Michael Engel}
\email{michael.engel@fau.de}
\affiliation{Institute for Multiscale Simulation, IZNF, Friedrich-Alexander University Erlangen-N{\"u}rnberg, 91058 Erlangen, Germany}

\date{\today}

\begin{abstract}
	Polyhedral nanocrystals are building blocks for nanostructured materials that find applications in catalysis and plasmonics.
	Synthesis efforts and self-assembly experiments have been assisted by computer simulations that predict phase equilibra.
	Most current simulations employ Monte Carlo methods, which generate stochastic dynamics.
	Collective and correlated configuration updates are alternatives that promise higher computational efficiency and generate trajectories with realistic dynamics.
	One such alternative involves event-chain updates and has recently been proposed for spherical particles.
	In this contribution, we develop and apply event-chain Monte Carlo for hard convex polyhedra.
	Our simulation makes use of an improved computational geometry algorithm XenoSweep, which predicts sweep collision in a particularly simple way.
	We implement Newtonian event chains in the open source general-purpose particle simulation toolkit HOOMD-blue for serial and parallel simulation.
	The speed-up over state-of-the-art Monte Carlo is between a factor of 10 for nearly spherical polyhedra and a factor of 2 for highly aspherical polyhedra.
	Finally, we validate the Newtonian event-chain algorithm by applying it to a current research problem, the multi-step nucleation of two classes of hard polyhedra.
\end{abstract}
	
\maketitle

\includegraphics[width=\linewidth]{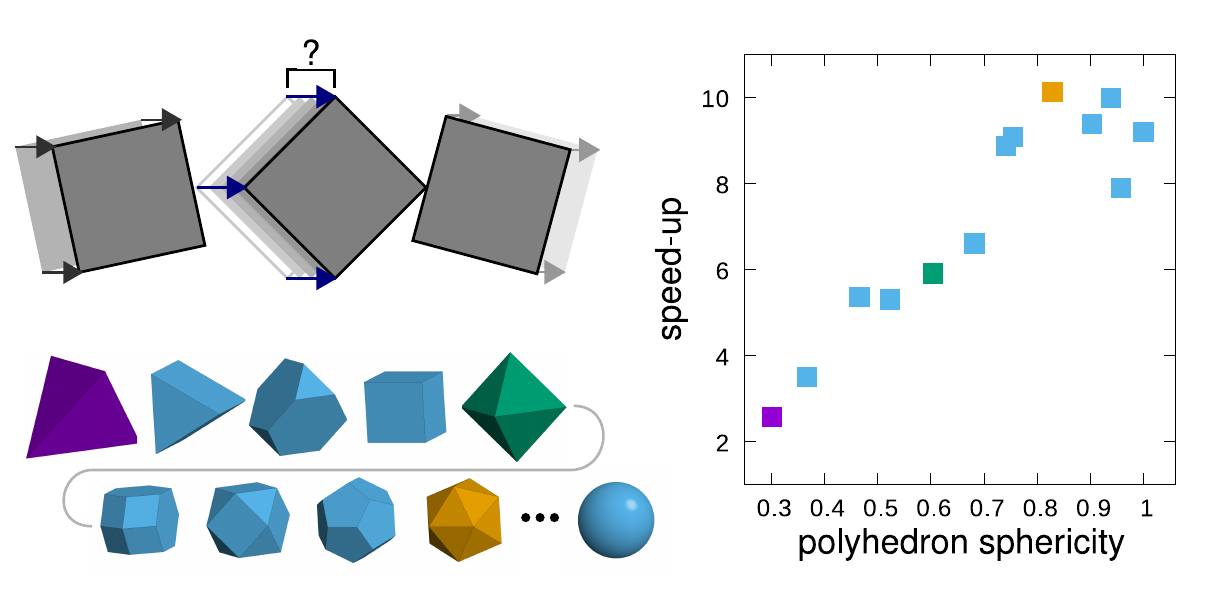}

\section{Introduction}

Nanocrystals can be synthesized in a variety of polyhedral Wulff shapes.
They self-assemble by destabilization in solution or in evaporation and sedimentation experiments\cite{chang_hydrothermal_2008,eguchi_simple_2012,henzie_self-assembly_2012,avci_self-assembly_2017}.
The resulting superlattices have applications as functional materials.\cite{boles_self-assembly_2016}
Some particles have several competing superlattice candidates.
Octahedra, for example, can form four different superlattice structures.
Three of these are extremal in packing density or contact area between particles.\cite{chang_hydrothermal_2008,gong_shape-dependent_2017}
While the structures predicted can be reproduced in experiment reliably, the ordering phenomena and pathways to the superlattices are often not well understood.
Varying the types and amounts of ligand molecules affects the structures found.\cite{wang_cuboctahedral_2016,lu_unusual_2019}
Only for special cases, predictive rules have been found.
One such case are hard particles, in which shape alone is the dominating factor.
Hard particle phase diagrams\cite{ni_phase_2012,damasceno_crystalline_2012,damasceno_predictive_2012,gantapara_phase_2013} and equations of states\cite{marechal_density_2013,irrgang_virial_2017,tian_equations_2019} help to understand experiments and to select promising polyhedral shapes and target structures.
Other studies utilize simulations outcomes to predict photonic properties.\cite{cersonsky_pressure-tunable_2018}

Simulations of anisotropic particles are often performed stochastically with Monte Carlo methods.
Molecular dynamics would in principle be favored because it equilibrates configurations more efficiently\cite{klement_efficient_2019}.
The reason is that density fluctuations relax with the speed of sound in molecular dynamics, whereas such fluctuations equilibrate by diffusion in Monte Carlo.
Furthermore, molecular dynamics creates realistic trajectories, which are an advantage when comparing details of trajectories with experiment.
However, integrating the equations of motion for anisotropic particles is not easily possible.
This is in particular the case for hard anisotropic particles, which interact via discontinuous forces.
Event-driven molecular dynamics is a well-established method for hard spheres to obtain Newtonian trajectories\cite{alder_studies_1960}.
But event-driven molecular dynamics for anisotropic particles requires solving collision equations that involve rotations and thus trigonometric functions.
The solution of these non-linear equations is only possible numerically by iteration with approximations\cite{donev_neighbor_2005}
and is necessarily slow.
In Monte Carlo, the problematic collision checks are replaced by overlap checks.
Importantly, overlaps can be solved analytically in a finite number of steps for many shapes, including for polyhedra.\cite{canny_collision_1986,gilbert_fast_1988,gilbert_fast_1988,snethen_xenocollide:_2008}
It is thus desirable to combine the advantages of molecular dynamics (fast equilibration and realistic dynamics) with the advantages of Monte Carlo (simple algorithm and no approximation).

Here, we develop a simulation method that combines the advantages of Monte Carlo and molecular dynamics in an algorithm that avoids approximations and iterations to the extend possible.
Our starting point is event-chain Monte Carlo\cite{bernard_event-chain_2009,kampmann_parallelized_2015}, which offers efficient equilibration and prediction of structures, especially in the recently proposed variant of Newtonian event chains.\cite{klement_efficient_2019}
We generalize Newtonian event-chain Monte Carlo to hard convex polyhedra.
The algorithmic bottleneck for event-chain Monte Carlo is the prediction of collisions between non-rotating polyhedra.
The detail that the polyhedra are non-rotating is crucial.
It means the underlying equations do not contain trigonometric functions and can be solved analytically.

Fast collision detection is a classic problem in computer graphics\cite{jimenez_3d_2001,brochu_efficient_2012,huang_survey_2019}.
Solutions started with the evaluation of all vertex--face and edge--edge feature combinations of the two polyhedra in question.\cite{canny_collision_1986}
Because the number of such combinations increases rapidly with the number of polyhedron vertices, a brute-force calculation of all combinations is too time consumption.
The algorithm by Gilbert, Johnson, and Kerthi\cite{gilbert_fast_1988} (GJK) was a major breakthrough.
It was not only faster, but supported all sorts of convex objects -- as long as there is a support function that returns the furthest vector of an object for any given direction.
Due to its conceptual complexity, the GJK algorithm remains cumbersome to implement.
Snethen\cite{snethen_xenocollide:_2008} created with Xenocollide a derived algorithm that can easily be visualized at any step.
Xenocollide is the starting point of the collision prediction algorithm presented in this paper.
Similar to the directional contact range\cite{choi_determining_2010} calculation, we determine the directional contact distance of convex polyhedra.
This is the distance by which a polyhedron can be translated without rotation up to collision, a process called `sweeping'.
The algorithm we present here, called XenoSweep, predicts sweep collisions and is particularly simple.

The central goal of the present manuscript is method development.
We first implement Newtonian event-chain Monte Carlo of polyhedral particles utilizing XenoSweep in an open source software package and test it extensively.
We then confirm that the algorithm indeed has the expected efficiency advantages.
An advantage of Newtonian event-chain Monte Carlo is that the partially stochastic trajectories generated resemble Newtonian dynamics significantly better than the fully stochastic trajectories of Monte Carlo with local moves only.

\section{Monte Carlo Simulation}

We consider a system of anisotropic particles.
The system is fully specified by the position vectors $\{\vec x_i\}_i$ and orientation quaternions $\{\vec q_i\}_i$ of all particles and their shape.
The most commonly employed strategy to simulate anisotropic particles is Monte Carlo with local updates of the positions and orientations.

\subsection{Local Updates}

Local-update Monte Carlo (LMC) applies a sequence of trial moves to update the configuration.
A randomly selected particle is either translated by a random vector of length up to $d_{\text{trans}}$ or rotated by addition of a random quaternion\cite{frenkel_understanding_2001} of norm up to $d_{\text{rot}}$ and subsequent re-normalization.

We denote the probability to execute a translation move as
\begin{equation}
	p_\text{trans} = \frac{\sharp \text{ translation trials}}{\sharp \text{ translation trials} + \sharp \text{ rotation trials}},
\end{equation}
where the character `$\sharp$' means `number of'.
The new configuration is accepted with a Boltzmann factor.
The Boltzmann factor collapses to an overlap check when simulating hard particles.

\subsection{Newtonian Event Chains}

Event-chain Monte Carlo (ECMC)\cite{bernard_event-chain_2009} applies a sequence of collective moves in the form of chains to update the configuration.
Each chain consists of a random start followed by a series of deterministic collision events.
A randomly selected particle is translated in a random chain direction up to its first collision.
The collision partner then takes over and continues the translation in the chain direction up to its first collision.
At this point the third particle takes over and so on.
ECMC terminates the chain after a predefined chain length.

\begin{figure} 
	\includegraphics[width=0.7\columnwidth]{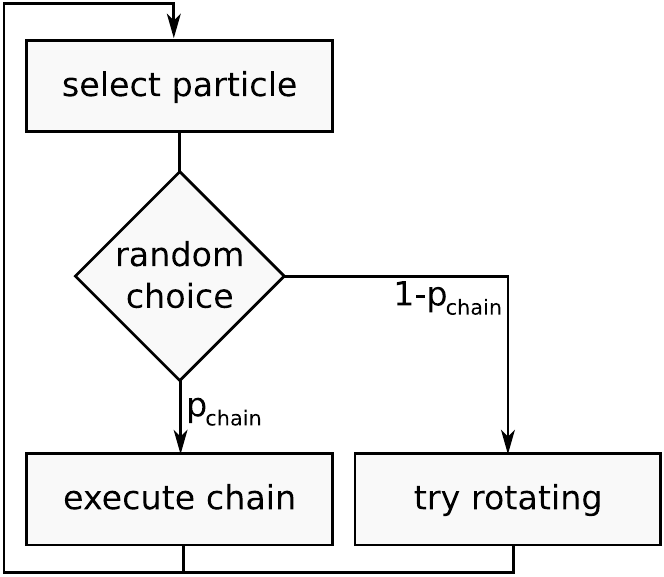}
	\caption{Flowchart of Monte Carlo simulation for anisotropic particles.
		The algorithm combines event chains for translation moves and random trial rotation moves as in local Monte Carlo.
		The type of move is chosen randomly.
		$p_\text{chain}$ is the probability to execute an event chain.
	}
	\label{fig.flowchart}
\end{figure}

ECMC only translates particles and has been applied only to disks and spheres\cite{bernard_event-chain_2009}, and point particles with interaction potentials.
To treat anisotropic particles, we mix event-chain translations with LMC rotation trial moves as illustrated in \cref{fig.flowchart}.
The probability to execute an event chain is
\begin{equation}
	p_\text{chain} = \frac{\sharp \text{ event chains}}{\sharp \text{ event chains} + \sharp \text{ rotation trials}}.
\end{equation}

We recently introduced Newtonian event-chain Monte Carlo (NEC) as a variant of ECMC that is more efficient but still follows the correct statistics, which is guaranteed by the fact that it obeys the balance condition.\cite{klement_efficient_2019}
NEC associates each particle with a velocity vector.
A randomly selected particle is translated in the direction of its velocity vector up to the first collision.
During this translation, time advances by $\Delta t_i = \Delta x_i / v_i$, where $\Delta x_i$ is the translation distance and $v_i$ the velocity of the translated particle.
The velocities of both particles involved in the collision are updated at the collision according to the rules of elastic collisions.
The chain then continues with the collision partner as in ECMC.
NEC terminates the chain after a predefined chain time $t_\text{chain}=\sum_i \Delta t_i$.

\subsection{Parameters}

LMC and NEC both have three dimensionless parameters, which must be tuned for maximal efficiency of the simulation algorithms:

(1)~The NEC chain time is naturally expressed in units of the mean free time $t_\text{mf}=\langle\Delta t_i\rangle_\text{NEC}$.
The average is conveniently computed in NEC after equilibration.
In LMC, the translation trial move distance is naturally expressed in units of the mean free path $d_\text{mf}$.
We approximate the mean free path using the mean free time and the root mean square velocity measured in NEC,
$d_\text{mf} \approx t_\text{mf} \sqrt{\langle v^2 \rangle}$.
We call the joint parameter the translation parameter. It is given by
\begin{equation}
	\tau=
	\begin{cases}
		\dfrac{d_{\text{trans}}}{d_\text{mf}} & \text{for LMC},\\
		\dfrac{t_\text{chain}}{t_\text{mf}} & \text{for NEC}.
	\end{cases}
\end{equation}
The translation parameter $\tau$ behaves differently in LMC and NEC and will be treated separately below.

(2)~The rotation trial move distance $d_{\text{rot}}$ is identical in LMC and NEC.
We call this parameter the rotation parameter,
\begin{equation}
	\rho=
	\begin{cases}
		d_{\text{rot}} & \text{for LMC},\\
		d_{\text{rot}} & \text{for NEC}.
	\end{cases}
\end{equation}

(3)~The probability to execute a translation move $p_\text{trans}$ in LMC can be mapped to an equivalent NEC parameter using $p_\text{chain}$ by observing that the number of translation moves is given by $p_\text{chain} (N_\text{chain}+1)$, where $N_\text{chain}$ is the average number of collision events per chain.
We call the fraction of translation moves the move ratio parameter.
It has the expression
\begin{align}
	\mu&= \frac{\sharp \text{ translation moves}}{\sharp \text{ translation moves} + \sharp \text{ rotation moves}} \nonumber\\
	&=\begin{cases}
		p_\text{trans} & \text{for LMC},\\
		\dfrac{p_\text{chain} (N_\text{chain}+1)}{p_\text{chain} (N_\text{chain}+1) + (1-p_\text{chain})} & \text{for NEC}.
	\end{cases}
\end{align}

\subsection{Parallel Event Chains}

ECMC can be parallelized with a cell decomposition scheme.\cite{kampmann_parallelized_2015}
As cells have a large inactive volume, we use domain decomposition\cite{anderson_scalable_2016} to parallelize NEC.
Collisions with the domain wall and with particles outside of the cell are treated as elastic collisions with partners of infinite mass.
Such collisions fulfill the detailed balance condition and lead to ergodic dynamics if the domain walls are shifted every now and then.\cite{kampmann_parallelized_2015}

We argued that NEC is efficient because the trajectories it generates follow Newtonian dynamics more closely.\cite{klement_efficient_2019}
However, while collision between particles conserve momentum, collisions with domain walls or outside particles break momentum conservation and break Newtonian dynamics near the walls.
The result is a decrease of the advantage of NEC over LMC.
This decrease can be critical especially for small domains.

Anisotropic particles have an additional issue.
The number of chains per cell between two domain wall shifts is not fixed but Bernoulli-distributed with mean proportional to $p_\text{chain} N_\text{domain}$ and variance proportional to $p_\text{chain} (1-p_\text{chain}) N_\text{domain}$.
Here, $N_\text{domain}$ is the average number of particles in the domain.
Whenever domain walls are to be shifted, the simulations for all domains must wait for the slowest domain (with most chains).
The variance can be reduced by using shorter chains, increasing $p_\text{chain}$, or generally simulating longer between shifting domain walls.
Too short chains destroy the advantage of event chains and are not efficient.~\cite{klement_efficient_2019}

\section{Sweep Collisions of Convex Polyhedra}

Event chains require the prediction of sweep collisions.
A sweep collision is the collision of two particles when one of them is translated along a given direction, the other is not translated, and neither of them is rotated.
While predicting sweep collisions of spheres is trivial, no simple expression exists for anisotropic particles.

We develop an iterative algorithm for sweep collision prediction of convex polyhedra.
For this purpose, we simplify and extend Snethen's XenoCollide algorithm\cite{snethen_xenocollide:_2008}.
XenoCollide utilizes Minkowski portal refinement.
Portals are triangles that hit the origin when translated (or swept) along a normalized ray direction $\vec r$.
This means the portal ray, which is the line $\{\lambda \vec r \mid \lambda \in \mathbbm{R}\}$, intersects the triangle.
Our modified XenoSweep algorithm not only returns the sweep distance $\ell$ at collision but also the normal vector to the collision plane $\vec n$.
The normal vector is not calculated by XenoCollide, but we need it to perform elastic collisions in ECMC or NEC.

Let $A$ and $B$ be two convex particles.
Minkowski portal refinement utilizes the Minkowski difference $C = \{\vec a - \vec b \mid \vec a\in A, \vec b\in B\}$.
$C$ contains the origin, if and only if the intersection of $A$ and $B$ is non-empty.
Sweep collision prediction is equivalent to checking whether $C$ is a portal. 
A further simplification is that we need to know only the support vector $S_C(\vec v)=\vec x$, which is defined as the vector $\vec x\in C$ that maximizes $\vec x\cdot\vec v$.
The support vector corresponds to the point in $C$ that is extended furthest along $\vec v$.

\begin{figure}
	\includegraphics[width=\columnwidth]{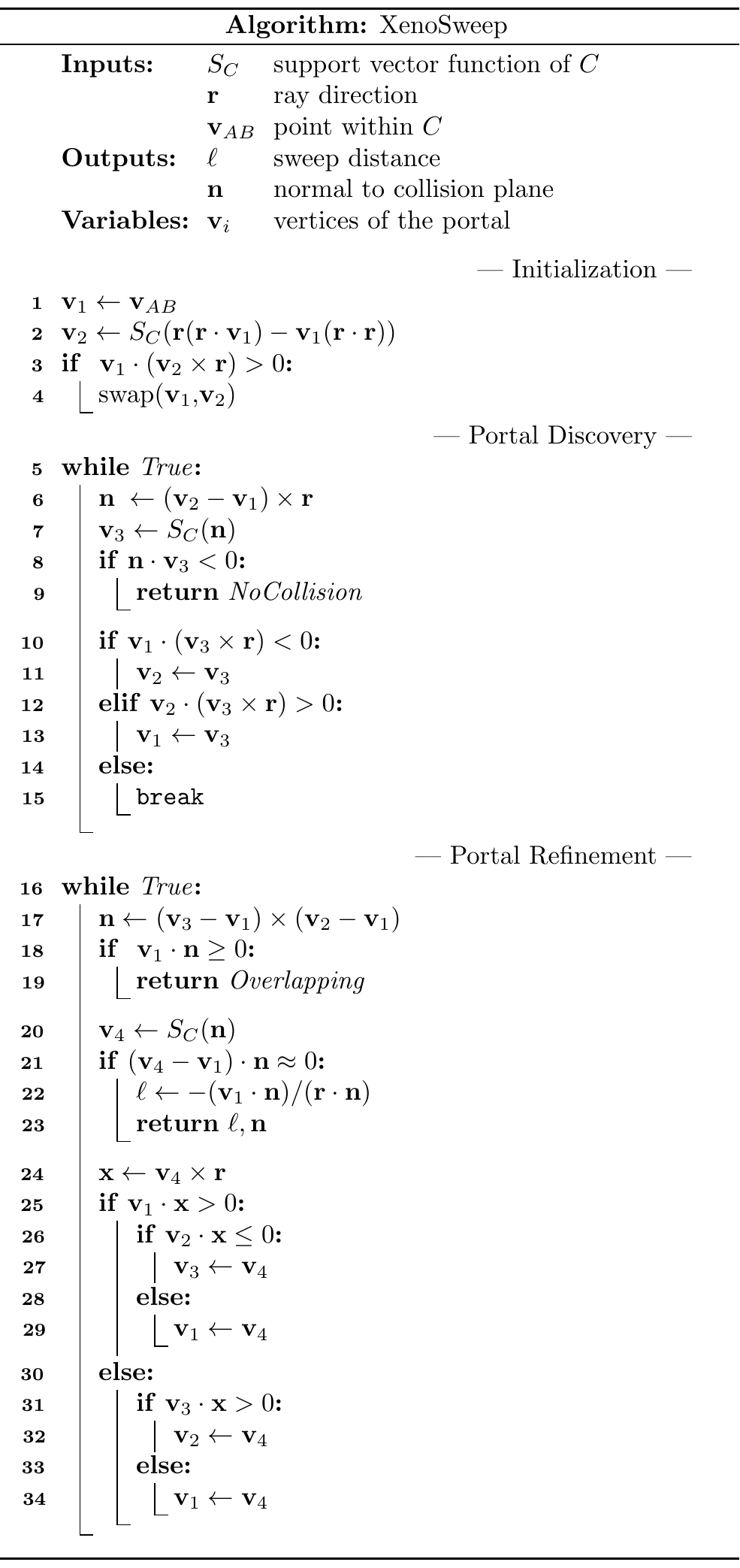}
	\caption{
		Python-like pseudocode of XenoSweep that extends XenoCollide\cite{snethen_xenocollide:_2008}.
		The algorithm consists of the three stages: Initialization, Portal Discovery, and Portal Refinement.
	}
	\label{fig.sweep}
\end{figure}

The XenoSweep algorithm (\cref{fig.sweep}) consists of three stages, an initialization stage and two iterative stages.
Both iteration stages terminate for polyhedra after a finite number of iterations bounded from above by the number of vertices.
In principle, the algorithm can be made to scale significantly faster than linearly with the number of polyhedron vertices by implementing a tree-like data structure for the calculation of the support vector function.
However, we do not implement such a data structure here to keep the algorithm simple.
The first iterative stage searches for the existence of a portal only in the plane perpendicular to the ray direction $\vec r$.
The second iterative stage gradually improves the portal towards the collision plane.
Differences to XenoCollide are a modified initialization and improved exit conditions.

\subsection{Initialization}
We initialize the algorithm with two vectors  $\vec v_1, \vec v_2\in C$.
These vectors can be chosen arbitrarily, which is what we do for $\vec v_1$ (line 1).
We find it efficient to set $\vec v_2$ as the support vector in direction $\vec r \times (\vec r \times \vec v_1)=\vec r(\vec r\cdot \vec v_1) - \vec v_1(\vec r\cdot\vec r)$ on the opposite side of the particle as seen along $\vec r$ towards the origin (line 2).
The normal vector $\vec n = \vec r \times (\vec v_2 - \vec v_1)$ is the search direction in the next step.
We guarantee that $\vec n$ points towards the origin, by swapping the initial vectors if that is not yet the case by testing the condition $\vec v_1 \cdot \vec n < 0$ (lines 3-4).

\subsection{Portal Discovery}
\begin{figure}
	\includegraphics[width=\columnwidth]{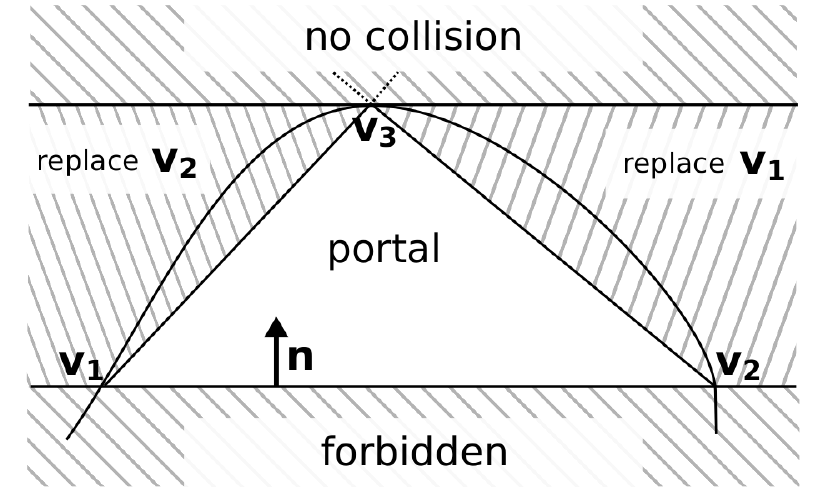}
	\caption{Sketch of the portal discovery stage.
		The portal candidate $(\vec v_1, \vec v_2, \vec v_3)$ is shown in projection along $\vec r$.
		The portal ray is a point in this projection.
		By construction, we know this point is located above the line through $\vec v_1$ and $\vec v_2$ and falls within the portal candidate if and only if the portal candidate is a portal.
		Portal discovery is iterated in the algorithm until either a portal is found or it is determined that no portal exists.}
	\label{fig.sketch2D}
\end{figure}

The portal discovery stage searches for the existence of a portal.
A portal candidate $(\vec v_1, \vec v_2, \vec v_3)$ is created by including as $\vec v_3$ the support vector in a direction perpendicular to $\vec r$ and towards the origin (lines 6-7).
By construction, we can exclude that the origin is behind the portal in negative normal direction (`forbidden' region in \cref{fig.sketch2D}).
If $\vec n \cdot \vec v_3 < 0$, we found a separating axis perpendicular to the ray direction $\vec r$ and can return `NoCollision' (lines 8-9).

Otherwise, we test the intersection of the portal ray with the portal candidate.
Depending on the relative location of the portal ray, portal discovery continues by replacing the vector furthest away (regions `replace $\vec v_1$ and `replace $\vec v_2$' in \cref{fig.sketch2D}), or breaks and continues with the next stage if the portal candidate is found to be a portal (lines 10-15).

\subsection{Portal Refinement}
At this point in the algorithm, we successfully discovered the existence of a portal ($\vec v_1$, $\vec v_2$, $\vec v_3$).
But other portals might be closer to the origin.
The distance of the portal from the origin along $\vec r$ is $\ell = -(\vec n\cdot\vec v_1)/(\vec n\cdot\vec r)$ with the portal normal vector $\vec n = (\vec v_3 - \vec v_1) \times (\vec v_2 - \vec v_1)$.
Here, we utilized that $\vec r$ is a normalized vector.
The portal refinement stage searches for the portal with the smallest $\ell$.
By construction $\vec n\cdot\vec r\ge0$, and we know $\ell\le0$ if $\vec n\cdot\vec v_1\ge0$, which means the origin is on the back-side of the portal.
At this point we cannot distinguish between particles moving apart and intersecting particles.
Because we are not interested in negative sweeps, we return `Overlapping' (lines 17-19).
If $\vec n\cdot\vec v_1\ge0$ first occurs in a later iteration, then the origin is located between the previous and the current portal.
Because all portals intersect $C$, the origin is then also inside $C$ and we know the particles are intersecting.

\begin{figure}
	\includegraphics[width=0.55\columnwidth]{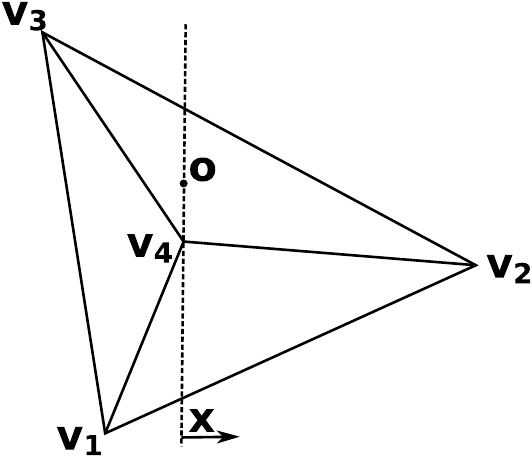}
	\caption{Sketch of the portal refinement stage.
		The portal $(\vec v_1, \vec v_2, \vec v_3)$ is projected along $\vec r$.
		We illustrate an example, where the origin $\vec O$ is located such that $\vec v_4$ replaces $\vec v_1$.}
	\label{fig.sketch3D}
\end{figure}

Next, we compute a new support vector $\vec v_4$ in direction of the normal vector (line 20).
We skip the check for a separating axis, $\vec n \cdot\vec v_4 < 0$, that would allow an early return in case the goal was only an overlap check instead of calculating the sweep distance.
If $\vec v_4$ is coplanar to the other vectors (up to numerical precision), the portal is not moving towards the origin anymore, and we return $\ell$ as the sweep distance (lines 21-23).
Finally, the portal is updated by replacing one of its vertices by $\vec v_4$ such that the new triangle remains a portal.
We can replace $\vec v_{i}$, $i \in \{1,2,3\}$ by $ \vec v_4$ when $\vec v_{(i+1)\bmod 3}\cdot\vec x > 0$ and $\vec v_{(i+2)\bmod 3}\cdot\vec x \le 0$, where $\vec x = \vec v_4 \times\vec r$ (lines 24-34).
One example is illustrated in \cref{fig.sketch3D}.
Based on the result of a first scalar product, we can choose the second scalar product either such that one of the replacement criterion is fulfilled directly or such that both scalar products have the same sign.
Because the origin ray intersects the portal, the set of scalar products $\vec v_i \cdot \vec x$, $i=1,2,3$ contain at least one positive and one negative result, and one replacement criterion is fulfilled implicitly.

\subsection{Test and validation}

\begin{figure}
	\includegraphics[width=\columnwidth]{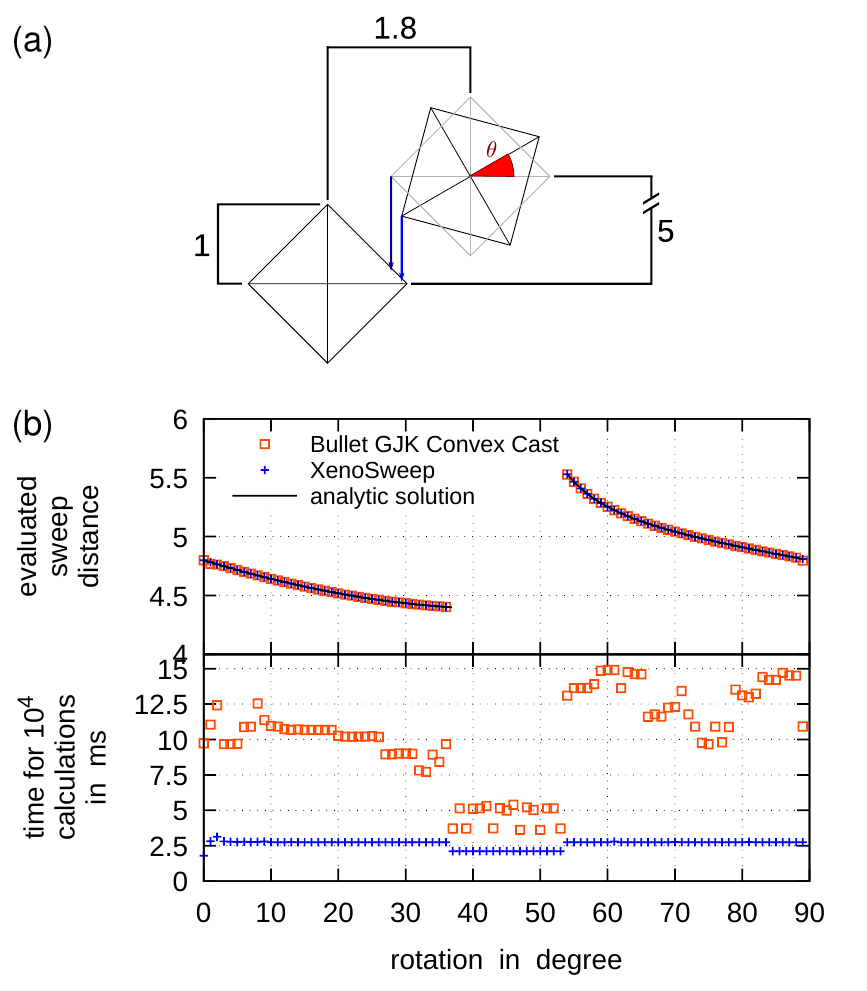}
	\caption{Comparison of collision prediction with XenoSweep and with the Bullet library.\cite{noauthor_bullet_nodate}
		(a)~Geometry for collision prediction tests.
		Two octahedra are placed with vertical and horizontal (not to scale) offset.
		(b)~The sweep distance of the octahedra is predicted accurately with both algorithms~(top).
		Time for the calculation of $10^4$ collisions on an Intel Core i5-2400 CPU~(bottom).
		In this test, XenoSweep is noticeably faster.
		The definition of the rotation angle $\theta$ (x-axis) is indicated in (a).
		}
	\label{fig.xeno.test}
\end{figure}

To validate XenoSweep, we set up a well-defined test configuration.
Two octahedra with edge length $1$ are displaced by $1.8$ in horizontal direction and $5$ in vertical direction (\cref{fig.xeno.test}(a)).
The sweep distance is calculated in vertical direction as the second octahedron is rotated.
For some rotation angles around $\pi/4$, the sweep distance is infinite because the octahedra do not collide.
An advantage of this setup is that the sweep collision can be calculated analytically.

\cref{fig.xeno.test}(b) compares the accuracy and performance of XenoSweep with \texttt{LinearConvexCastDemo} in the Bullet library \cite{noauthor_bullet_nodate}, which includes collision detection by employing GJK Ray Casting\cite{van_den_bergen_ray_2004}.
The Bullet library is a physics engine and is used in many video games and for visual effects in movies.
Our calculations show that the evaluated sweep distance is accurate in both algorithms and agrees with the analytic solution.
We also evaluate the time for $10^4$ collision.
In this test, XenoSweep is significantly faster with a performance that only weakly depends on the specific geometry.
We find an improvement of about a factor of 4 in cases with collision, and an improvement of about a factor of 2 if there is no collision.
While this single test cannot replace a systematic benchmark, it already demonstrates that XenoSweep is not only simple but also sufficiently fast for our purposes, and that it compares well with established collision detection libraries.
A further, more thorough validation will be presented below where we utilize NEC for the reproduction of multi-step nucleation in two thermodynamic systems of hard polyhedra.

\section{Implementation and Parameterization}

We implemented NECs in the open source general-purpose particle simulation toolkit HOOMD-blue.\cite{anderson_general_2008,anderson_scalable_2016}
HOOME-blue has a well-tested and highly-efficient hard particle Monte Carlo (HPMC) package, which we extend.
From the existing HOOMD-blue codebase, we use axis-aligned bounding boxes (AABBs), AABB-trees, the general management structure for memory, communication via MPI, and file input/output.
As a result of this work, NEC for spheres\cite{klement_efficient_2019} and convex polyhedron will be included in HOOMD-blue in version 3.0.0.\cite{noauthor_httphoomd-bluereadthedocsio_nodate}

\begin{figure}
	\includegraphics[width=0.8\columnwidth]{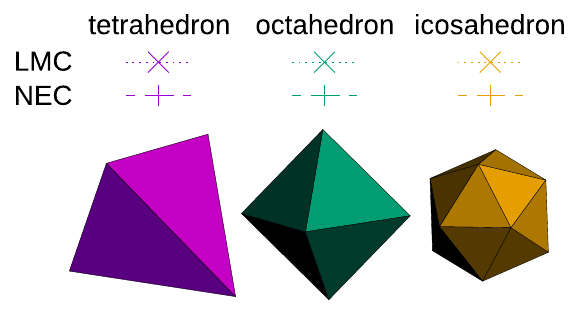}
	\caption{The three polyhedra used for optimizing the parameters of LMC and NEC.
	The colors and symbols represent the key for \cref{fig.par.rot,fig.par.trans,,fig.par.mr}.}
	\label{fig.key}
\end{figure}

It remains to tune the translation parameter, the rotation parameter, and the move ratio parameter in both LMC and NEC for maximal efficiency.
Three polyhedra are tested, the tetrahedron with four vertices, the octahedron with six vertices, and the icosahedron with twelve vertices (\cref{fig.key}).
Every polyhedron is scaled to have volume $V_0=1$.
The reference systems contains 8000 particles at 45\% volume fraction.
At this density all convex polyhedron remain in the fluid phase.
Particle velocities in NEC are initialized randomly with mean velocity $\langle\vec v\rangle = 0$ and root mean square velocity $\sqrt{\langle\vec v^2\rangle}=1$.
We quantify the efficiency of the algorithms by measuring the diffusion coefficient in units of CPU time (seconds), $D_\text{CPU}$.
Simulations are performed in single-core mode on Intel Skylake compute nodes.
Unless specified otherwise in each of the following sections, we use:
$\tau_\text{LMC}=2$ in LMC, $\tau_\text{NEC}=30$ in NEC, $\mu=0.5$,
$\rho=0.15$ for icosahedron and octahedron; $\rho=0.07$ for tetrahedron.
The unit of length $x$ in \cref{fig.par.rot,fig.par.trans,,fig.par.mr} is the length of a cube with the same volume for each shape.

\subsection{Translation parameter}

\begin{figure}
	\includegraphics[width=\columnwidth]{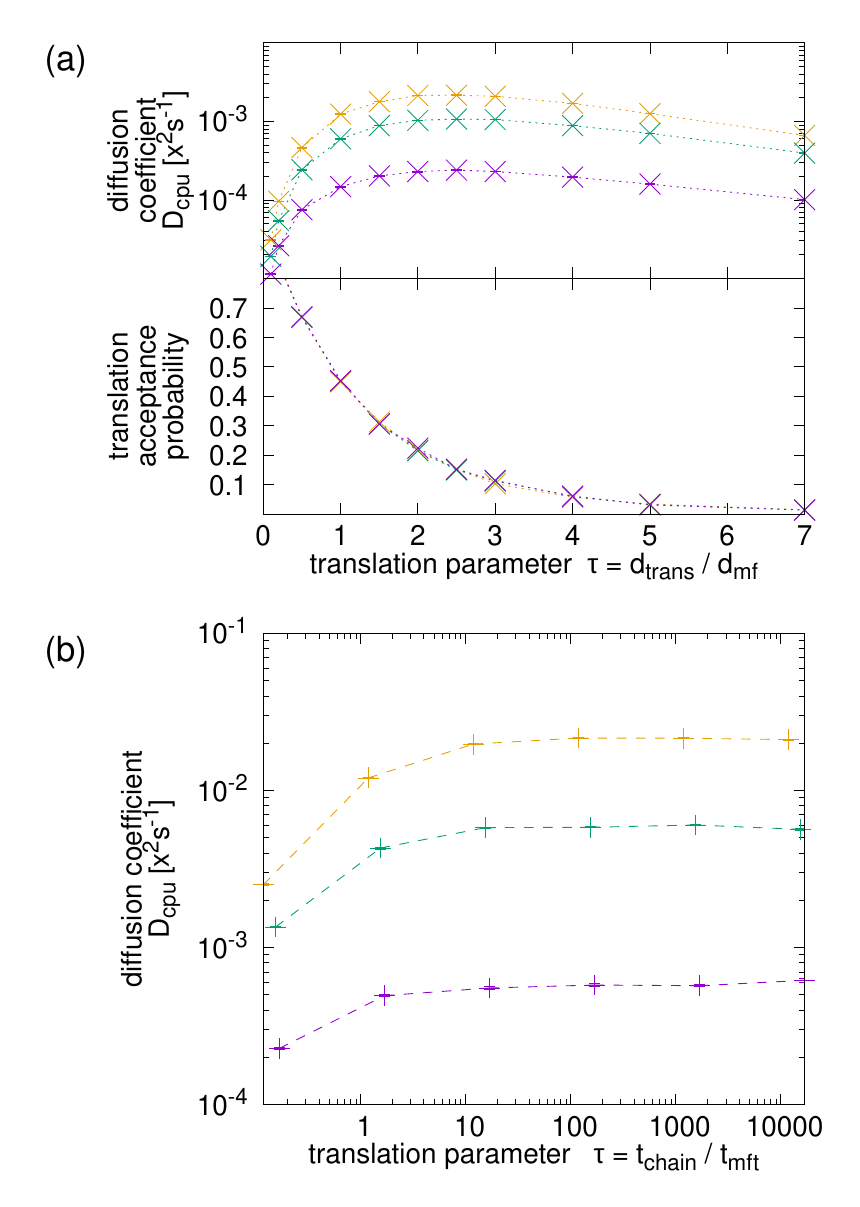}
	\caption{Effect of the translation parameter $\tau$ on the diffusion coefficient $D_\text{CPU}$ for (a)~LMC and (b)~NEC.
		Highest diffusion occurs near $\tau = 2$ with a corresponding LMC acceptance probability of about 20\%.
		As $\tau$ increases, diffusion decreases for LMC but plateaus for NEC.
		}
	\label{fig.par.trans}
\end{figure}

We discuss the translation parameter $\tau$ separately for LMC and NEC because it affects the efficienty of both algorithms differently.
In LMC, highest diffusion is obtained for a translation move distance of about double the mean free path, i.e.\ $\tau = 2$ (\cref{fig.par.trans}(a)).
The acceptance probability of translation moves is about $20\%$ for this choice.
Diffusion eventually decreases with increasing $\tau$ because random translations over large distances generate many overlaps.

In NEC, chains should be sufficiently long ($\tau \ge 10$) but increasing their length further has no significantly detrimental effect (\cref{fig.par.trans}(b)).
This behavior agrees with NEC of hard spheres\cite{klement_efficient_2019}.
In both cases, particles that are more spherical (icosahedron; yellow color) diffuse faster than particles that are less spherical (tetrahedron; purple) with the octahedron (green) located in the middle.
We also observe a clear performance advantage (higher diffusion) of NEC over LMC that is greater for the icosahedron than the tetrahedron.

\subsection{Rotation parameter}

\begin{figure}
\includegraphics[width=\columnwidth]{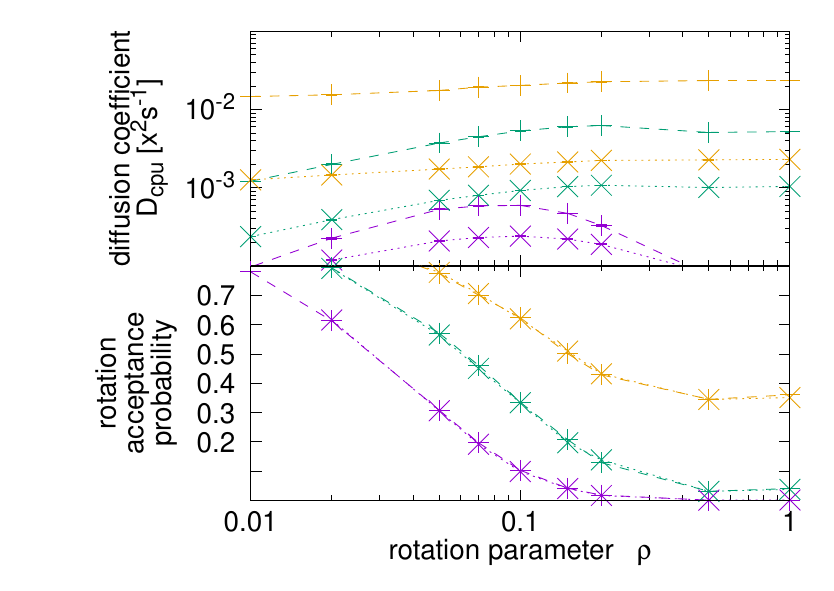}
\caption{Effect of the rotation parameter $\rho$ on the diffusion coefficient $D_\text{CPU}$.
	Highest diffusion occurs for acceptance probabilities near $20\%$.
	We expect diffusion to decrease faster along the $\rho$-axis in more dense systems.}
\label{fig.par.rot}
\end{figure}

The rotation parameter $\rho$ has similar behavior in LMC and NEC (\cref{fig.par.rot}).
At density $45\%$, particles still have ample space to rotate, which means optimal performance is obtained for rather large orientation changes.
Just like for the translation parameter in LMC, optimal performance generally occurs near values of the acceptance probability of around $20\%$.
An overall vertical shift of the curves can be explained by the role of translation updates, which dominate diffusion.
As before, more spherical particles diffuse faster, and NEC is generally more efficient than LMC.

\subsection{Move ratio parameter}
\begin{figure}
	\includegraphics[width=\columnwidth]{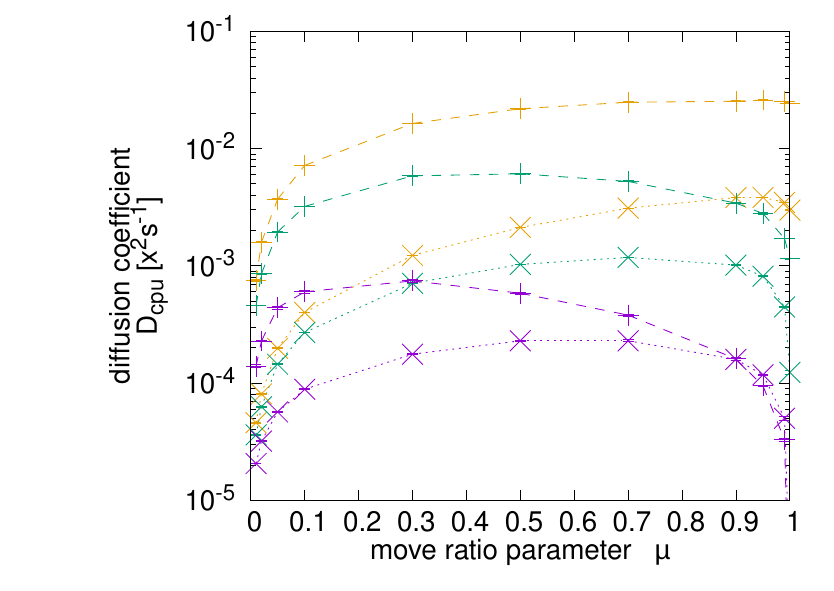}
	\caption{Effect of the move ratio parameter $\mu$ on the diffusion coefficient $D_\text{CPU}$.
		As expected, a mix of rotation and translation gives highest diffusion.
		The optimal mix changes with the polyhedron shape.
		}
	\label{fig.par.mr}
\end{figure}

The move ratio $\mu$ is the ratio of translation moves among all moves in the simulation and ranges from zero (only rotation) to one (only translation).
Simulations with parameters in either end of the parameter range are inefficient because particles get easily stuck in their neighbor shell (\cref{fig.par.mr})).
Shape plays an important role for the optimal choice of $\mu$.
Icosahedra prefer more translations ($\mu\gtrsim 0$), which is explained by the presence of a rotator phase prior to crystallization.\cite{mravlak_phase_2018}
Tetrahedra prefer more rotations ($\mu\lesssim1$) due to their strong preference for face-to-face contact.\cite{haji-akbari_disordered_2009}
In general, we expect nearly spherical polyhedra to rotate easily and highly anisotropic polyhedra to preferentially align face-to-face.

Generally, the location of the diffusion maximum shifts towards lower $\mu$ in NEC compared to LMC.
This confirms once more that translation moves are more efficient in NEC.
We do not see a clear choice for $\mu$ that would allow us to select a particular optimal parameter value.
For this reason, we set $\mu=0.5$ in line with common practice.

\section{Performance}

After extending the NEC algorithm to convex polyhedra and implementing it in HOOMD-blue, we now quantify the speed-up of NEC over LMC.
For this purpose, we analyze the ratio of the diffusion coefficients, $D_\text{CPU}^\text{NEC} / D_\text{CPU}^\text{LMC}$.
The simulation parameters of the algorithms are chosen as $\tau_\text{LMC}=2$, $\tau_\text{NEC}=30$, $\mu=0.5$.
The rotation parameter is chosen depending on particle shape to account for variations in the importance of rotations as $\rho=0.07$ for the tetrahedron, $\rho=0.10$ for the triangular prism, $\rho=0.15$ for the truncated tetrahedron, the cube, and the octahedron, $\rho=0.3$ for the elongated dodecahedron, and $\rho=1.0$ for all others polyhedra investigated.

We put a particular focus on the role of particle shape, which we describe by the sphericity (or isoperimetric quotient) of the polyhedron.
Sphericity is defined for a convex particle as $Q = 36\pi V^2 A^{-3}$ with surface area $A$ and volume $V$.
Besides single-core performance, we also investigate the scaling behavior for parallel LMC and parallel NEC.

\subsection{Single-Core Performance}

\begin{figure}
	\includegraphics[width=\columnwidth]{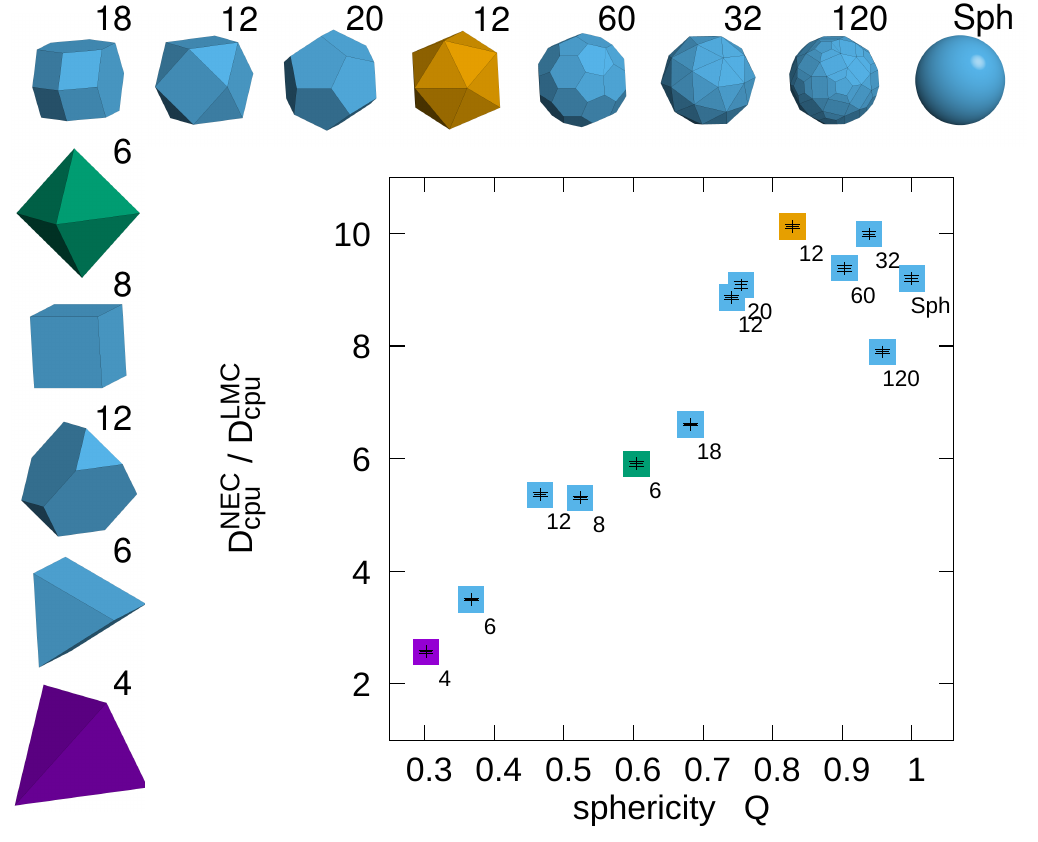}
	\caption{
		Speed-up of NEC over LMC as quantified by the ratio of diffusion coefficients.
		Twelve convex polyhedra and the sphere (labelled `Sph') are tested.
		The polyhedra are, from bottom left (number of vertices): tetrahedron~(4), triangular prism~(6), truncated tetrahedron~(12), cube~(8), octahedron~(6), elongated dodecahedron~(18), cuboctahedron~(12), dodecahedron~(20), icosahedron~(12), truncated icosahedron~(60), pentakis dodecahedron; all vertices of same distance from the origin~(32), and truncated pentakis dodecahedron~(120).
		}
	\label{fig.shapes}
\end{figure}

We analyze the speed-up of NEC over LMC for different polyhedra in serial (single-core) mode of HOOMD-blue.
As \cref{fig.shapes} shows, NEC is significantly more efficient than LMC for all tested convex polyhedra.
The speed-up is always at least a factor of 2 and can be up to a factor of 10, with the minimum taken for tetrahedra and the maximum for icosahedra.

We observe the trend that higher sphericity generally results in higher speed-up (\cref{fig.shapes}).
This makes sense because, as we have seen in \cref{fig.par.mr}, translations are less critical for equilibrating polyhedra with low sphericity.
After all, translations benefit directly from NEC whereas rotations are not or only indirectly affected by the efficiency boost.
The relationship between sphericity and speed-up is nearly linear.

The number of vertices also plays a role, but a minor one.
For example the pentakis dodecahedron (sphericity 0.939, 32 vertices) is sped up by a factor of 10 and the truncated pentakis dodecahedron (sphericity 0.958, 120 vertices) is sped up by a factor of 8.
The speed-up is smaller for the truncated pentakis dodecahedron, which has similar sphericity but more vertices than the pentakis dodecahedron.
This behavior can be explained by details of the XenoSweep algorithm.
While an overlap check with XenoCollide (implemented in LMC) only finds a separating axis or a shared point, the distance measurement with XenoSweep searches for the colliding surface elements.
The letter task costs more compute time and, in comparison, slows down faster with the number of polyhedron vertices.
Interestingly, the NEC speed-up for the sphere\cite{klement_efficient_2019} falls right on top of the polyhedron data despite the use of conceptually different algorithms for overlap and collision checks in both cases.

\subsection{Parallel Performance}

To test parallel performance of NEC, we simulate octahedra varying the system size and the number of CPU cores.
Parallel performance is analyzed by the particle update frequency (i.e., how fast the algorithm generates and executed translations and rotations) and the diffusion coefficient (i.e., the efficiency of the algorithm to propagate through configuration space).
We normalized the measurements to the values for a single-core simulation and divide them by the number of cores. 

\begin{figure}
	\includegraphics[width=\columnwidth]{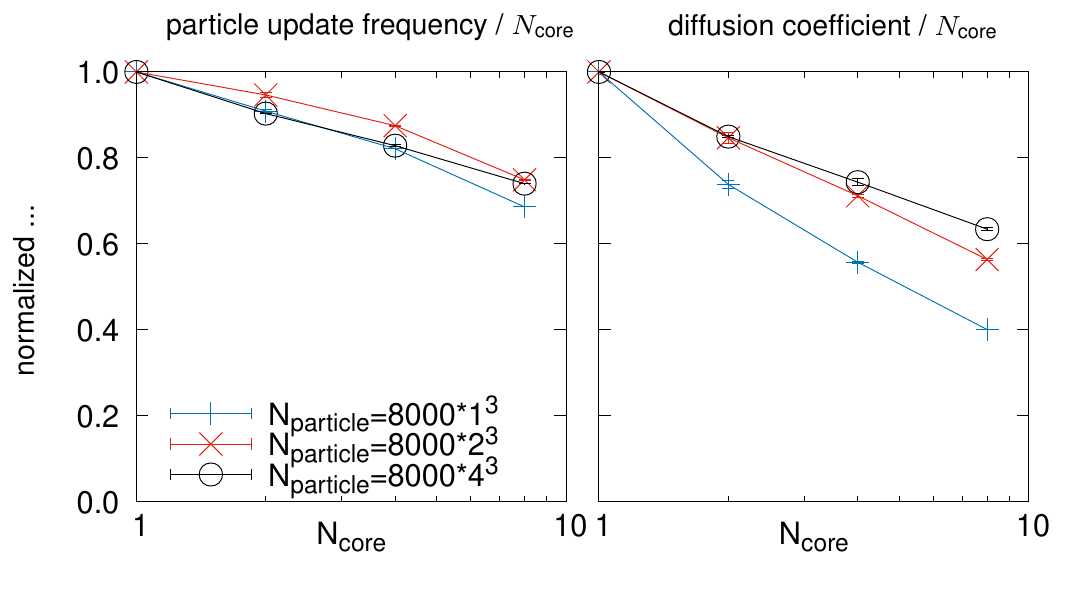}
	\caption{
		Scaling behavior for parallel NEC for systems of different numbers of particles $N_\text{particle}$.
		The algorithm ran on the same CPUs but utilized different numbers of cores $N_\text{core}\in\{1,2,4,8\}$.
		At all times, all CPU cores were completely filled with identical compute jobs to achieve comparable load.
		The particle update frequency (left) and the diffusion coefficient (right) are normalized to their single-core values and divided by the number of cores.
		A flat line corresponds to perfect scaling.
	}
	\label{fig.perf.parallel-log}
\end{figure}

Results are shown in \cref{fig.perf.parallel-log}.
NEC slows down as the number of cores increases, and scales better in the particle update frequency than the diffusion coefficient.
This means domain decomposition has a negative influence on the parallel performance of chains, an effect that we find to be particularly strong for small systems.
Our observation is in line with previous research.\cite{kampmann_parallelized_2015}
NEC requires correlated translations in chains over significantly larger distances than translation trial moves in LMC.
This can become problematic.
For good parallel efficiency of NEC beyond a small number of cores, the system should be rather large, likely contain millions of particles.
Massively parallel simulations on GPUs as performed with LMC\cite{anderson_massively_2013}
are not advisable with NEC, though chains with local times\cite{li_multithreaded_2021}
may be an alternative.

\section{Pathways of multi-step nucleation}

\begin{figure*}
	\centering
	\includegraphics[width=0.65\textwidth]{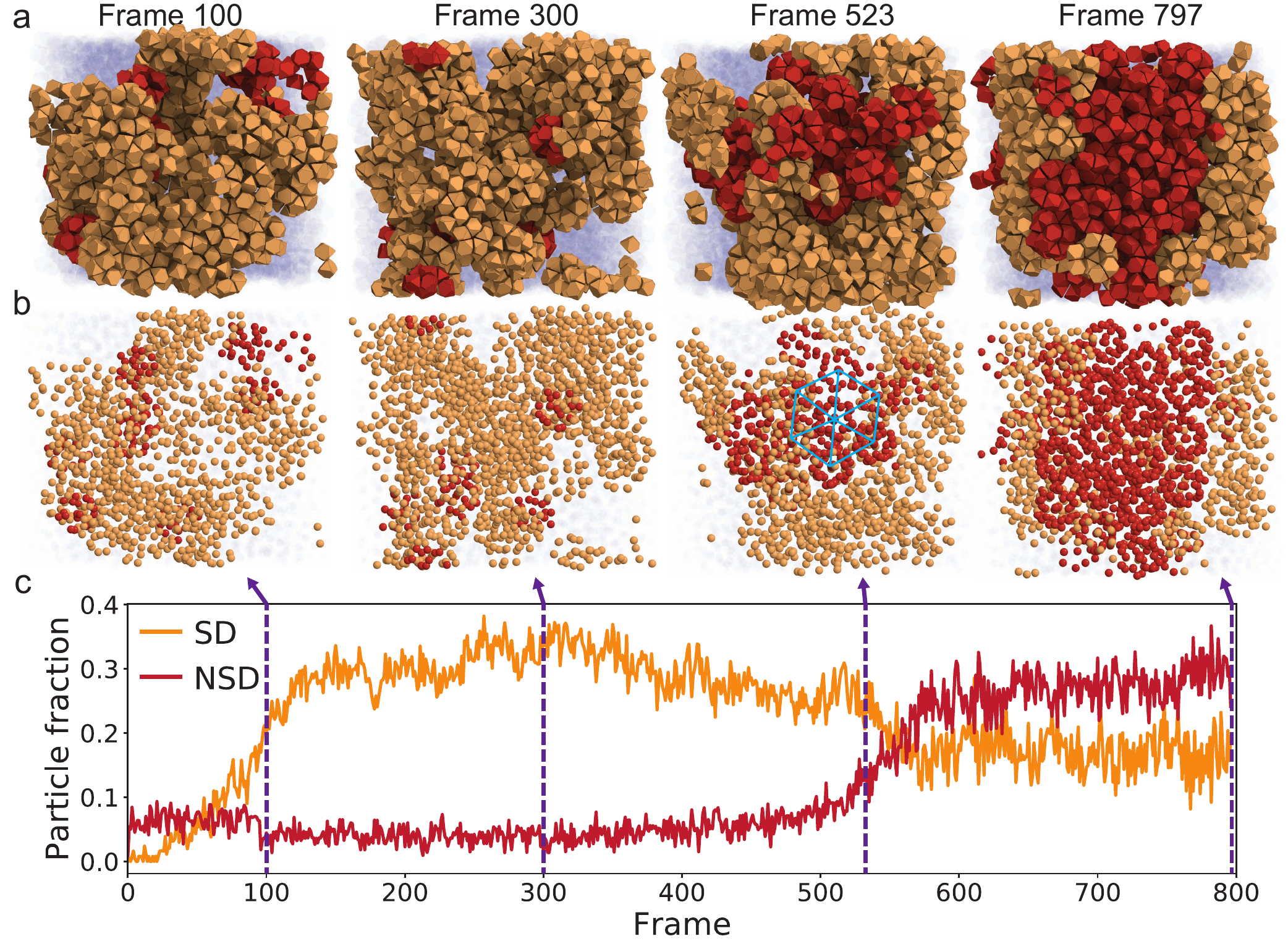}
	\caption{
		Analysis of a long NEC trajectory for tetrahedra with truncated tips and edges (TTs) at volume fraction $61\%$.
		The system is known to crystallize in two steps via a prenucleation motif with clusters.\cite{lee_entropic_2019}
		Views along the $(111)$ direction depicting (a)~the polyhedra and (b)~spheres at the centers of the polyhedra.
		Particles are colored for better visibility of the crystal, with fluid particles translucent.
		A high-density fluid phase (orange, left) features a dodecahedron motif with particles shared across dodecahedron (SD).
		The crystal phase (red, right) features dodecahedron that do not share particles (NSD).
		(c)~Characterization of the trajectory with order parameters.
		See Ref.~\citenum{lee_entropic_2019} for a detailed description of the phases and order parameters.
	}
	\label{fig.application.TTT}
\end{figure*}

\begin{figure*}
	\centering
	\includegraphics[width=0.65\textwidth]{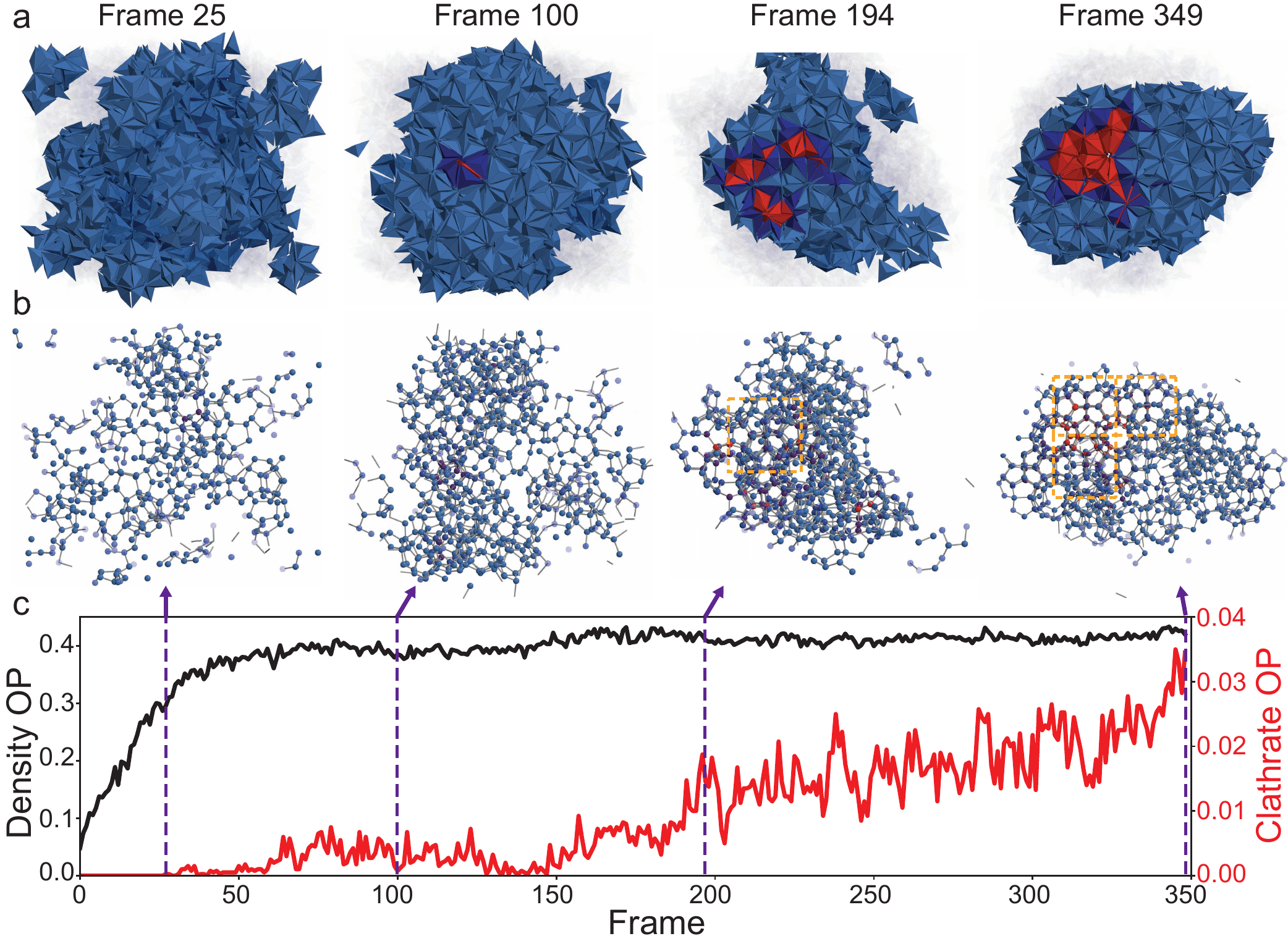}
	\caption{
		Analysis of a long NEC trajectory for triangular bipyramids (TBPs) at volume fraction $50\%$.
		The system is known to crystallize in two steps via a prenucleation motif with a dense network.\cite{lee_entropic_2019}
		Views along the $(100)$ direction depicting (a)~the polyhedra and
		(b)~spheres at the centers of the polyhedra tetramers.
		Particles are colored for better visibility of the crystal, with fluid particles translucent.
		A high-density fluid phase (blue) features tetrahedral coordination.
		The crystal phase (red) is of type clathrate I.
		(c)~Characterization of the trajectory with two order parameters.
		See Ref.~\citenum{lee_entropic_2019} for a detailed description of the phases and order parameters.
	}
	\label{fig.application.TBP}
\end{figure*}

It remains to apply the polyhedron NEC code to a current research problem.
We focus on systems of truncated tetrahedra (TTs) and triangular bipyramids (TBPs).
Both of these systems show interesting phase behavior and unusual phase transformation pathways.\cite{lee_entropic_2019}
TTs forms a high-density fluid consisting of shared dodecahedron motifs that nucleate a cF432 crystal.
TBPs form a high-density fluid in form of an amorphous network and crystallize into a Clathrate~I crystal.
Importantly, when simulated at intermediate volume fraction, both systems exhibit two-step crystallization.
This finding was unexpected and the systems are currently the only known examples of multi-step phase transitions in a hard particle system.
Our aim here is to test whether NEC can reproduce the findings previously made with the LMC implementation in the HPMC package of HOOMD-blue.
This is a good test, because such phase behavior is highly sensitive to mistakes in the collision prediction (i.e.\ problems in the XenoSweep algorithm) and to violations of the balance condition (i.e.\ incorrect statistics in the NEC implementation).
Our analysis utilizes the same order parameters as in previous work.\cite{lee_entropic_2019}

Indeed, TTs (\cref{fig.application.TTT}) and TBPs (\cref{fig.application.TBP}) both reproduce all details of the reported two-step crystallization pathways.
This finding demonstrates that our NEC algorithm follows the correct thermodynamics.
Furthermore, NEC simulations of 4000 particles reach initial crystal formation after 430 core-hours (0.11~h/particle) for TTs and 292 core-hours (0.07~h/particle) for TBPs.
In contrast, previous LMC simulations required 1725 core-hours (TT) for a system of 8000 TT particles (0.22~h/particle) and 1800 core-hours for a system of 20,000 TBP particles (0.09~h/particle) for crystallization to start.\cite{lee_entropic_2019}
This means our observation is a first indication that it is possible to convert the speed-up derived from brief trajectories and diffusion measurements into an efficient equilibration of highly complex particle systems with a noticeable (factor 2.0) speed advantage of NEC over LMC for TTs a slight (factor 1.2) speed advantage for the highly aspherical TBPs.
Still, given the difference in system sizes, these numbers are difficult to compare.
Furthermore, because crystallization is a stochastic process that becomes more probable with larger system size, these speed advantages might in fact be underestimations.
Clearly, more simulation work is necessary for a rigorous comparison.

As an added benefit, our simulations directly address a concern related to the way how Monte Carlo trajectories approach equilibrium.
By design, Monte Carlo and molecular dynamics methods must always reach the same equilibrium given sufficient equilibration time.
This equivalence of algorithms is, however, not necessarily the case for trajectories:
Monte Carlo trajectories are distinct from trajectories produced by molecular dynamics at the microscopic level.
Molecular dynamics trajectories are preferred, in principle, because they resemble the dynamics of experiments more closely.
But molecular dynamics is not easily achievable for polyhedra, because the equations that need to be solved are highly non-linear.
Event-driven hard particle simulations work only for spheres or with numerical approximations\cite{donev_neighbor_2005}
that might again affect the trajectories.
In contrast, Monte Carlo is much simpler and therefore currently preferentially applied to anisotropic particles.
A priori there is no guarantee that LMC reaches equilibrium in the same way as integrating Newton's equations of motions.
Is the use of Monte Carlo as a replacement for molecular dynamics problematic when analyzing details of phase transformations trajectories?

NEC does not fully reproduce Newtonian dynamics but resembles it significantly better than Monte Carlo\cite{klement_efficient_2019}.
In fact, NEC can reasonably be considered intermediate to Monte Carlo and molecular dynamics.
Our results in \cref{fig.application.TTT,fig.application.TBP} demonstrate that the phenomenon of two-step crystallization remain present in NEC.
This indicates that the choice of simulation algorithm is, after all, less crucial.
We note that these results are only a preliminary evaluation of the phenomenon.
More in-depth analysis will be required in future to settle the issue more conclusively.

\section{Conclusion}

We developed, implemented, and tested NEC for convex hard polyhedra.
The new simulation method is an improvement over existing polyhedron simulation codes.
It is multiple times more efficient, cutting simulation time and resulting in a lower cost and smaller CO$_2$ footprint.
As an added benefits, NEC returns particle trajectories that are closer to molecular dynamics than conventional Monte Carlo simulation.
Another advantage is that the virial expression for pressure, often a helpful observable to identify phase transformations, drops out as a side product of using event chains without additional effort
\cite{michel_generalized_2014,engel_hard-disk_2013,isobe_hard-sphere_2015,klement_efficient_2019}.
In LMC, pressure must be computed separately and requires additional steps.
Our method is a first step towards more general and more versatile simulation algorithms for anisotropic particles.

At the core of NEC lies the sweep collision prediction algorithm XenoSweep.
This algorithm heavily builds on prior work in the computer graphics community.\cite{snethen_xenocollide:_2008}
Our new implementation is an important improvement both in terms of speed and in terms of simplicity of the algorithm.
XenoSweep solves the classic problem of polyhedron collision prediction and overlap detection\cite{williams_discrete_1999,de_berg_computational_2008}
 in only 34 lines of pseudocode.
Such efficient and simple collision prediction is of interest to researchers working on a broad range of problems in computer graphics, robotics, and granular dynamics.

Future extensions should advance our work in two directions: improved handling of rotations and generalization to anisotropic particles with extended interaction, i.e.\ particles that are not purely hard.
Concerning the first point: While we sped up the effect of translation moves by about one order of magnitude, rotations remain a bottleneck.
This is apparent from \cref{fig.shapes}.
Efficiency reduces as the importance of rotations increases, which is the case towards more aspherical particles.
Future work could develop collective rotations moves that resemble conceptually the idea of event chains for translations.
Such an improved algorithm has the potential to reach a speed-up of up to one order of magnitude across all convex polyhedra.
Second, the hard particle condition employed exclusively in this work is a simplification and limits comparison to experiments.
Whereas nanoparticle shape is often of central importance, interactions such as van der Waals forces, electrostatic forces, or the effect of ligands can typically not be entirely ignored.
Developing a general simulation method for anisotropic convex or concave polyhedra with arbitrary interactions remains a challenging open problem.

\section*{Notes}

The authors declare no competing financial interest.
An implementation of the XenoSweep algorithm and NEC for convex polyhedra is included open-source as part of the HPMC package of the general-purpose particle simulation toolkit HOOMD-blue in version 3.0.0.\cite{anderson_general_2008,anderson_scalable_2016}

\begin{acknowledgments}
	This work has been funded by SFB1411 and the Cluster of Excellence Engineering of Advanced Materials of the German Research Foundation (DFG).
	Support by the Central Institute for Scientific Computing (ZISC), the Interdisciplinary Center for Functional Particle Systems (FPS), and computational resources and support provided by the Erlangen Regional Computing Center (RRZE) are gratefully acknowledged.
	Algorithm implementation and optimization in HOOMD-blue supported by the National Science Foundation, Division of Materials Research Award $\#$ DMR 1808342. 
	Work implementing the test cases and model verification was supported as part of the Center for Bio-Inspired Energy Science, an Energy Frontier Research Center funded by the U.S. Department of Energy, Office of Science, Basic Energy Sciences under Award $\#$ DE-SC0000989.
	This work used the Extreme Science and Engineering Discovery Environment (XSEDE)\cite{Towns2014}, which is supported by National Science Foundation grant number ACI-1548562; XSEDE award DMR 140129.
\end{acknowledgments}

\section*{Bibliography}

%

\end{document}